\documentstyle[prl,twocolumn,aps]{revtex}
\input{epsf}

\def\tr{{\rm tr}\,} 
\def\Tr{{\rm Tr}\,} 
\newcommand{\beq}{\begin{equation}}
\newcommand{\eeq}{\end{equation}} \newcommand{\beqn}{\begin{eqnarray}}
\newcommand{\eeqn}{\end{eqnarray}}

\newcommand{\diag}{{\rm diag}}

\begin{document}
\draft


\twocolumn[\hsize\textwidth\columnwidth\hsize\csname @twocolumnfalse\endcsname
\


\title{ Chern-Simons Theories on Noncommutative Plane
}\vspace*{-0.22cm}
 
\author{Dongsu Bak,${}^a$
Kimyeong Lee,${}^b$ and Jeong-Hyuck Park${}^b$}

\address {Physics Department, 
University of Seoul, Seoul 130-743 Korea${}^a$}
\address {Korea Institute for Advanced 
Study, 207-43, Cheongryangri-dong, Dongdaemun-gu, Seoul 
130-012, Korea ${}^b$}

\date{\today}

\maketitle
\begin{abstract}
%
 We investigate $U(N)$ Chern-Simons theories on noncommutative plane.
We show that for the theories to be consistent quantum mechanically,
the coefficient of the Chern-Simons term should be quantized $\kappa =
n/2\pi$ with an integer $n$. This is a surprise for the $U(1)$ gauge
theory. When  uniform background charge density $\rho_e$ is present, the
quantization rule changes to $\kappa +\rho_e\theta = n/2\pi$ with 
noncommutative parameter $\theta$.  With the exact expression for the
angular momentum, we argue in the $U(1)$ theory that charged particles
in the symmetric phase carry fractional spin $1/2n$ and vortices in
the broken phase carry half-integer or integer spin $-n/2$. 

\end{abstract}
]

\newpage

Recently there has been a considerable interest in the noncommutative
gauge theories with Yang-Mills kinetic term~\cite{noncomm}. The Wilson
loop line segments and the covariant position operators play a crucial
role in defining the covariant semi-local
densities~\cite{wilson,wess,jhp}.  The energy momentum tensors are
gauge covariant and so the issue of the gauge invariant local operator
has been discussed in many directions.

The Chern-Simons gauge theories on noncommutative plane have been
studied also recently~\cite{chen,khare,dongsu}. There was a claim that
there is no need for the Chern-Simons coefficient in the
noncommutative case~\cite{shj}.

In this letter, we show that the Chern-Simons coefficient should be
quantized if the noncommutative $U(N)$ Chern-Simons theories are 
consistent quantum mechanically.  This is true even for the $U(1)$
theory.  One can also introduce an additional uniform charge as the
background, in which case only a linear combination of the background
charge and the Chern-Simons coefficient is quantized. We also find the
conserved angular momentum and calculate it for any rotationally
symmetric configuration. The result is identical to the commutative
case, implying that the spin-statistics theorem holds even for the
noncommutative field theories.

Recently Susskind~\cite{susskind} argued that the noncommutative
version of the $U(1)$ Chern-Simons theory at level $n$ is exactly
equivalent to the Laughlin theory of the filling fraction $1/n$. In
this argument, the quantization of the Chern-Simons coefficient is
essential. The present paper  provides the proof of the quantization.

The noncommutative  plane is defined by the coordinates
$(x,y)$ satisfying the commutative relation
\beq [x,y]= i\theta. 
\label{comm}
\eeq
We choose $\theta>0$ without loss of generality.  The Hilbert space of
harmonic oscillator is defined by the annihilation and creation
operators
\beq
a = \frac{x+iy}{\sqrt{2\theta}},\;\;~~~~~~~ \bar{a} =
\frac{x-iy}{\sqrt{2\theta}}
\eeq
which satisfy $[a,a^\dagger]=1$.  The space integration $\int d^2x $
becomes $2\pi \theta {\rm Tr}$. 

For the $U(N)$ gauge group, the gauge field $A(x)$ is hermitian matrix
valued operator on the noncommutative space. For simplicity, we just include
a charged matter field $\phi(x)$. The action for the theory is the sum of the
Yang-Mills action, the Chern-Simons action, and the matter action. 
The Yang-Mills action is
\beq
S_{YM}  =     \int dt \; 2\pi \theta \; \Tr  \left( -\frac{1}{4e^2}\tr
F_{\mu\nu}F^{\mu\nu} \right) \,,
\eeq
where $F_{\mu\nu} = \partial_\mu  A_\nu -\partial_\nu A_\mu
-i[A_\mu,A_\nu]$.  The small $\tr$ is over the $N$ by $N$ group
indices. The Chern-Simons action~\cite{roman} is 
\beq
S_{CS} = \!\!\! \int dt2\pi \theta \Tr \!    \frac{\kappa}{2} 
\epsilon^{\mu\nu\rho} \tr \biggl( A_\mu 
\partial_\nu A_\rho -\frac{2i}{3} A_\mu A_\nu A_\rho \biggr) . 
\eeq
The matter action is
\beq
S_{matter} = \int dt\; 2\pi \theta \Tr \tr \left( -\nabla_\mu \phi
\nabla^\mu \bar{\phi} - U(\phi,\bar{\phi}) \right) \,, 
\eeq
where the covariant derivative is $\nabla_\mu \phi = (\partial_\mu
-iA_\mu) \phi$.

Under the local $U(N)$ gauge transformation $g= e^{i\Lambda(x)}$, the
gauge field $A_\mu$ transforms as $ 
A_\mu(x) \rightarrow g A_\mu \bar{g} -i\partial_\mu g \bar{g}, $ 
and $\phi \rightarrow g \phi$. The Yang-Mills and matter actions are
manifestly invariant under this transformation. However, the
Chern-Simons action is not manifestly invariant. Rather its change is
\beq
\Delta  S_{CS} =  i\pi\kappa \theta   \int dt\; \Tr \tr
\epsilon^{\mu\nu\rho} \partial_\mu ( 
\partial_\nu   g  A_\rho \bar{g}) -4\pi^2  \kappa {\cal N}\,, 
\label{gtr}
\eeq
where 
\beq 
{\cal N} =   \frac{1}{24 \pi^2 } \int dt\;  2\pi \theta \Tr\tr
\epsilon^{\mu\nu\rho}  \left(  \bar{g}  \partial_\mu g \bar{g}
\partial_\nu g \bar{g} \partial_\rho g \right)\,.
\eeq
This quantity ${\cal N}$ is the analogue of the `topological number'
for the mapping from $S^3 \rightarrow SU(2)$. 
As  $\partial_i g = [\hat{\partial}_i, g]$ with $
\hat{\partial}_i = \frac{i}{\theta} \epsilon_{ij} x^j $, 
the topological number becomes
\beq
{\cal N}(g) 
= -\frac{i}{4\pi} \int dt\; \Tr \tr \left( \bar{g}
\partial_t g \bigl[ \bar{g} [a,g], \bar{g}[\bar{a},g]\bigr] \right)\,. 
\eeq
As we make successive gauge transformations which go to the identity
quickly at infinity, their topological numbers add up. Thus, we see
that ${\cal N}(g_1 g_2) = {\cal N}(g_1) +{\cal N}(g_2)$.

The gauge transformation is $g(x) = e^{i\Lambda(x)}$ with a hermitian
matrix valued function $\Lambda (x)$. As $\Lambda$ is hermitian also
as an operator on the Hilbert space, we can find a unitary
transformation $U(x)$ which diagonalize $\Lambda$ and so $g$. That is
$g = U e^{i\Lambda_D} \bar{U}$ with a diagonal $\Lambda_D$.  The
topological charge would not change with this diagonalization. Thus we
need to calculate the topological number for $g=e^{i\Lambda_D}$.
The general expression for the diagonal function $\Lambda_D(x)$ is
\beq
\Lambda_D (x) = \sum_l \diag \biggl( f_{1l}(t), f_{2l}(t),..., f_{Nl}(t)
\biggr)
|l\rangle \langle l|\,,
\eeq
where $f_{al}(t)$ are the real functions of time only.  The contribution
of each diagonal element is  independent and adds to the topological
charge, 
\beq
{\cal N}(g)  =
\frac{1}{2\pi} \sum_{la} \left(f_{al}(\infty)- f_{al}(-\infty) \right)\,.
\eeq
For gauge transformations which go to the identity transformation at
the spacetime infinity, $f_{al}(t)$ approaches $2\pi \times ({\rm
integer})$ as $t \rightarrow \pm \infty$ and $f_{al}(t)\rightarrow 0 $
as $l\rightarrow \infty$.  Thus the above topological charge should be
integer  even in the noncommutative $U(1)$ gauge theory.

The path integral amplitude for a given configuration  is proportional to
$ {\cal A} = e^{i S_{CS}}\,, $
which can change nontrivially under the gauge transformations which
go to the identity at spacetime infinity. That change should be
unity for the theory to be consistent, 
\beq
e^{i\Delta S_{CS}} = e^{ -i4\pi^2 \kappa {\cal N}} = 1\,,
\eeq
for arbitrary integer ${\cal N}$. This consistency  leads to the
quantization rule 
\beq
\kappa = \frac{n}{2\pi}\, 
\label{quant}
\eeq
with an integer $n$. Thus the Chern-Simons coefficient should be
quantized for  the  noncommutative $U(N)$ gauge theory.  The parameter $n$ is
also called the `level' of the Chern-Simons term. This quantization rule
is identical to that for the commutative theory with larger gauge  group but
is novel for the noncommutative  $U(1)$ theory. 

It seems that the gauge transformations considered above do not have
the smooth limit when $\theta\rightarrow 0$. One can write the
noncommutative version of  more familiar map from $R^3 \rightarrow
SU(2)$ with nonzero wrapping. Define a generator of $U(N)$ group
\beq
\Theta(z,\bar{z})  = \bar{W}(\bar{z}) \frac{1}{W\bar{W}} W(z) \,
\label{Theta}
\eeq
with holomorphic $N$ dimensional row vector $W(z)$ whose components
are polynomials of order $l$. Consider a gauge transformation
$g=e^{-if(t) \Theta(\infty)} e^{i f(t) \Theta(z,\bar{z})}$ with the
boundary condition that $f(\pm\infty) = 2\pi m_{\pm}$ with integers
$m_\pm$. This gauge transformation goes to the identity at spacetime
boundary. This gauge transformation has the topological charge ${\cal
N} = -(m_+-m_-) l$ and also a smooth $\theta=0$ limit. The expression 
(\ref{Theta}) has a smooth $\theta=0$ limit except for  the $U(1)$
theory. (When
$W$ is antiholomorphic, one gets opposite topological number if
$W\bar{W}$ is invertible.) A similar expression for the above gauge
transformation appears also in the calculation of the topological
number for noncommutative $CP(N)$ theories~\cite{yang}.

A different  form of the Chern-Simons action  can be obtained by
using the covariant position operator
\beq
X^i = x^i +\theta \epsilon_{ij} A_j\,. 
\label{covp}
\eeq
These operators appear in the matrix interpretation of the
noncommutative Yang-Mills term.  The Chern-Simons term becomes
\beq 
S_{CS}' = \int dt\; 2\pi \theta  \Tr \left\{ \frac{\kappa}{2
\theta^2} \tr \epsilon_{ij} \nabla_0 X^i X^j -\frac{\kappa}{\theta}
\tr A_0 \right\}\,,
\label{alt}
\eeq
where $\nabla_0 X^i = \partial_0X^i -i[ A_0, X^i]$.  This expression
is possible only for the noncommutative gauge theory.  The first term
of the right side is manifestly gauge invariant. The second term is
not gauge invariant. Under the local gauge transformation
$e^{i\Lambda_D}$, it changes identically as before, leading to the
same quantization condition.  The quantization of the coefficient of
$A_0$ has appeared before.  In the gauged dynamics of quantum
mechanics with zero spatial dimension, the term linear in $A_0$ is a
one-dimensional analogue of the three dimensional Chern-Simons term
and the quantization of its coefficient in the $U(1)$ theory has been
explored before~\cite{dunne}. The reasons for the quantization in both
cases are identical.

This gauge noninvariant term has been studied also in different
context. The term proportional to $A_0$ is the background electric
charge density.  As explored in the Maxwell and Chern-Simons theories
before~\cite{piljin}, one can add an action which describes the effect
of uniform background electric charge density. In the Chern-Simons theory, it
can be translated to a uniform background magnetic field. The action
is
\beq
S_{back} = \int dt\; 2\pi\theta \Tr \left\{ -\rho_e \tr A_0 \right\}\,,
\eeq
where $\rho_e$ is constant. This action is not gauge invariant, and
looks identical to the last term for the alternative Chern-Simons
action in the previous paragraph.  Thus the gauge invariance implies
the modification of the quantization rule (\ref{quant}) to
\beq
\kappa + \rho_e \theta = \frac{n}{2\pi}\,.
\eeq
Only a linear combination of these two parameters is quantized.

The Chern-Simons theories can be used to describe anyons, or particles
of fractional spin and statistics. The commutation relation
(\ref{comm}) which defines the noncommutativity does not violate the
translation and rotation symmetries. Thus, we expect that there exist
corresponding conserved quantities. (See Ref.~\cite{jhp,dongsu} for the
previous discussion for these conserved quantities.) Under the
rotation $x^i\rightarrow x'^i = x^i + \delta x^i$ with $\delta x^i =
\epsilon_{ij} x^j$, the fields in the commutative theories transform
as $\delta \phi(x) = -\delta x^i \partial_i \phi $ and $ \delta A_i(x)
=\epsilon_{ij} A_j - \delta x^j \partial_j A_i $.  Here we have
dropped the infinitesimal parameter in the front of the variation.  We
can decompose the transformation of the fields into the gauge
covariant one plus a pure gauge transformation~\cite{angular}. In the
noncommutative field theories, one should be careful about the
ordering of coordinates. Especially, with $\phi'(x')=\phi(x)$ under
the rotation, we notice that in the fourier expansion
\beqn
\delta \phi &=& \int d^2 p\; \phi(p) \left( e^{i(x^i-\delta x^i) p_i} - 
e^{ix^i p_i} \right) \nonumber \\
&=&  -\frac{1}{2} (\delta x^i \partial_i \phi + \partial_i \phi \delta x^i )\,.
\eeqn
Here we have used the identity $e^{i(x^i-\delta x^i ) p_i } = e^{-i
\delta x^i p_i/2} e^{ix^ip_i} e^{-i \delta x^i p_i /2}$ when $\delta
x^i$ is linear in $x^i$.  Thus, the generalization of the
transformation to the  noncommutative case is that the transformation
$\delta \phi$ and $\delta A_i$ is the sum of the covariant
transformation
\beqn
&& \delta_{cov} \phi = \frac{1}{2} \epsilon_{ij} \left( X^i \nabla_j
\phi + \nabla_j \phi x^i  \right)\,, \nonumber  \\
&& \delta_{cov} A_i = -\frac{1}{2} (X^i F_{12} + F_{12} X^i)\,, 
\eeqn
plus a local gauge transformation $\delta_{gauge} \phi = i\Lambda \phi
$ and $\delta_{gauge} A_i = \nabla_i \Lambda$ with
\beq
\Lambda = \frac{1}{2} \left[ \epsilon_{ij}(x^i A_j + A_j x^i) + \theta
A_i^2 \right]\,.
\eeq
The above expression has the well defined commutative limit as $\theta
\rightarrow 0$.

The Noether charge we are interested in is that for the covariant
transformation $\delta_{cov} = \delta - \delta_{gauge}$. While the
action is invariant under $\delta$, the action under $\delta_{gauge}$
changes by  a total time derivative term after the Gauss law is
used. This is due to the Chern-Simons term and the
background charge.  The gauge invariant angular momentum is then
\beqn
J = && - 2\pi \theta \Tr \tr  \left\{ \frac{1}{e^2} F_{0i} \delta_{cov} A_i +
\nabla_0\phi \delta_{cov} \bar{\phi} + \delta_{cov} \phi \nabla_0 \bar{\phi} 
\right. \nonumber \\
&& \left. +\frac{\rho_e}{2} (X^i)^2 F_{12} \right\}\,.
\eeqn
We have discarded gauge noninvariant boundary terms, which are related
to the manifestly conserved current.  This is the generalization of
the angular momentum~\cite{piljin} to the noncommutative space. In the
symmetric phase $F_{12}$ can have a uniform nonzero value in the
symmetric phase, in which case this ground value should be substracted from
the above expression. (See for detail in Ref.~\cite{piljin}.)

To calculate the angular momentum of the various objects in the
theory, let us focus on the $U(1)$ gauge theory without the Maxwell
term nor the background charge. The angular momentum then has only the
matter contribution. We are interested in finding the explicit angular
momentum for an initial configuration of the gauge field and the Higgs
field. We assume that the initial velocity or time derivative is zero,
but there is nonzero $A_0$ to satisfy the Gauss law constraint. In
addition we assume that the initial configuration is rotationally
symmetric. The most general ansatz with the covariant position
operators instead of the gauge field is then
\beqn
&& \phi(x) = \sum_{l =0}^\infty f_l |l  \rangle \langle l  +\mu |\,,
\;\;\;\; A_0(x) = \sum_{l =0}^\infty a_l |l\rangle\langle l|\,,
\nonumber \\
&&  Z = X + iY = \sum_{l=0}^{\infty} \sqrt{2\theta (l +1 + k_l)}\; 
|l\rangle\langle l+1|\,, 
\eeqn
with real coefficients $f_l, a_l, k_l$. We call $\mu$ the vorticity of
the Higgs field at origin. This can be positive or negative
integer. In the negative case, the coefficients $f_0,.., f_{|\mu|-1}$
vanish. When $A_i = 0$, the coefficients $k_l$ vanish.  For localized
configurations the coefficient $k_l$ should converge to a finite value
in the large $l$ limit. Also we choose $k_{-1}=0$. As $ F_{ 12} =
-\frac{1}{\theta} + \frac{[Z,\bar{Z}]}{2\theta^2} $, the total
magnetic flux $\Psi = 2\pi\theta \Tr F_{12} $ becomes
\beq
\Psi = 2\pi \lim_{l \rightarrow \infty} k_l \equiv
2\pi (\mu + \alpha)\,.
\eeq
This limit can be positive or negative. 
Its analogue in the commutative theory is the expression for the
magnetic flux as the line integration of the gauge field at the
spatial infinity, $\Psi = \oint_{\infty} dl^i A_i $. 
The Gauss law $\kappa F_{12} -\rho = 0 $ where the charge density is
$\rho = - i(\nabla_0 \phi \bar{\phi} - \phi \nabla_0 \bar{\phi})$ becomes
\beq
\frac{\kappa}{\theta} (k_l-k_{l-1}) = -2 f_l^2 a_l\,.
\eeq
for all $l$.  The conserved electric charge $Q = 2\pi \theta \Tr \rho
= \kappa \Psi $ is
\beq
Q = 2\pi \kappa ( \mu+ \alpha )\,.
\eeq
In the symmetric phase, the electric charge is quantized in integer in
quantum mechanics as the field is invariant under $2\pi$ phase
rotation. A configuration without vorticity describes a collection of
elementary particles. A single particle would have the values $\mu=0$
and $\alpha = 1/(2\pi \kappa) = 1/n$, and so its magnetic flux is
fractional.

The angular momentum in the theory without the Maxwell term can be
rewritten as
\beqn
J = -&& i \pi  \Tr \left\{  (X^i)^2 (\nabla_0 \phi \bar{\phi} -
\phi \nabla_0 
\bar{\phi } ) \right. \nonumber \\
&& \left.  -  ( \nabla_0 \phi \,\, ( x^i)^2 \bar{\phi} - \phi \,\, (x^i)^2
\nabla_0\bar{\phi} ) \right\}\,.
\eeqn
The angular momentum for the rotationally symmetric ansatz becomes
\beq
J =   \pi \kappa (\alpha^2-\mu^2)\,.
\eeq
This expression is identical to one appeared in Ref.~\cite{yoonbai}.
When this result is combined with the quantization rule, the
consequence is fascinating.

In the symmetric phase, our initial configuration is a generic one which may
describe a q-ball with vortices in the middle. Many elementary
particles could make q-balls. Elementary particle has charge
quantized. In noncommutative theory, the charged matter field can
carry only unit charge. Thus, the magnetic flux is fractional
$1/\kappa=2\pi/n$.  The spin of elementary particles and antiparticles
with $\alpha=\pm 1/2\pi \kappa  $ and $\mu=0$ is identical to
\beq
s_{particle} =  \frac{1}{2n}\,, 
\eeq
which has a fractional spin as $n$ is an integer.  In the commutative
case, their statistics comes from the Aharonov-Bohm phase.  We expect
that is true even in the noncommutative case.

In the broken phase, the scalar field has nonzero expectation
value, $\lim_{l\rightarrow \infty } f_l = v $. For finite energy
configurations, the gradient energy $\nabla _i \phi
\nabla_i \bar{\phi}$ should vanish quickly, which implies that
$k_l\rightarrow \mu$ or $\alpha =0$. Then the magnetic flux is
determined by the vorticity $\Psi = 2\pi \mu$ as in the commutative
case. Elementary vortices and antivortices with $\alpha=0, \mu=1$ carry
the spin  
\beq
s_{vortex} = -\frac{n}{2}\,.
\eeq
Thus vortices in the broken phase can be fermions, bosons. The sign
difference between the spin of elementary particles and that of
vortices appears also in the commutative case. The
statistics of vortices is expected to be due to the Aharonov-Bohm
phase plus the Magnus phase, as in the commutative case~\cite{yoonbai}.

We have shown that the Chern-Simons coefficient is quantized on
noncommutative space. When the background charge is introduced, only a
linear combination of the Chern-Simons coefficient is
quantized. Furthermore, we argued that elementary particles in the
symmetric phase can have fractional spin but vortices in the broken
phase can be only fermions or bosons.


\noindent{\large\bf Acknowledgment} 

This work is supported in part by KOSEF 1998
Interdisciplinary Research Grant 98-07-02-07-01-5 (DB and KL) and by
UOS Academic Research Grant (DB). 

\noindent{\it Note Added}: Up on finishing this paper, a related
paper~\cite{nair} appeared in the hep-th.


\begin{thebibliography}{99}


\bibitem{noncomm}
A. Connes, M.R. Douglas and A. Schwartz, {\it Noncommutative
geometry and matrix compactification on tori}, JHEP 9802 (1992) 003,
hep-th/9711162; N. Seiberg and E. Witten, {\it String theory and
noncommutative geometry}, JHEP 09 (1999) 032, hep-th/9908142. 


\bibitem{wilson} N. Ishibashi, S. Iso, H. Kawai and Y. Kitazawa,
Nucl. Phys. B 573 (2000) 573, hep-th/9910004; J. Ambjorn,
Y.M. Makeenkom J. Nishimura and R.J. Szabo, {\it Lattice gauge fields
and discrete noncommutative Yang-Mills theory}, JHEP 0005 (2000) 023,
hep-th/0004147; D.J. Gross, A. Hashimoto and N. Itzhaki, {\it
Observables of non-commutative gauge theories}, hep-th/0008075.


\bibitem{wess} J. Madore, S. Schraml, P. Schupp and J. Wess,
Eur. Phys. J. C16 (2000) 161, hep-th/0001203.


\bibitem{jhp} D. Bak, K. Lee and J.-H. Park, {\it Comments on 
noncommutative gauge theories.}, hep-th/0011244.



\bibitem{chen} C-S. Chu. {\it Induced Chern-Simons and WZW action in
noncommutative spacetime}, Nucl. Phys. B580 (2000) 352,
hep-th/0003007; A.A. Bichl, J.M. Grimpstrup, V. Putz, and M Scheda,
{\it Perturbative Chern-Simons Theory on noncommutative $R^3$},
hep-th/0004071; G.-H. Chen and Y.-S. Wu, {\it One loop shift in
noncommutative Chern-Simons coupling}, hep-th/0006114;
A.P. Polychonakos,{\it Noncommutative Chern-Simons terms and the
noncommutative vacuum}, hep-th/0010264.; N. Grandi and G.A. Silva, {\it
Chern-Simons action in noncommutative space}, hep-th/0010113;
J. Kluson, {\it Matrix model and noncommutative Chern-Simons theory},
hep-th/0012184.


\bibitem{khare} G.S. Lozano, E.F. Moreno and F.A. Schaposnik,
{\it Self-dual Chern-Simons Solitons in noncommutative space},
hep-th/0012266; A. Khare and M.B. Paranjape, {\it Solitons in 2+1
dimensional noncommutative Maxwell-Chern-Simons Higgs theories},
hep-th/0102016.

\bibitem{dongsu}
D. Bak, S.K. Kim, K.-S. Soh and J.H. Yee, {\it Noncommutative
Chern-Simons Solitons}, hep-th/0102137.





\bibitem{shj}
M.M. Sheikh-Jabbari, {\it A note on noncommutative Chern-Simons
theories}, hep-th/0102092.



\bibitem{susskind}
L. Susskind, {\it The quantum Hall fluid and non-commutative
Chern-Simons theories}, hep-th/0101029.


\bibitem{roman}
S. Deser, R. Jackiw and S. Templeton, {\it Three dimensional
massive gauge theories}, Phys. Rev. Lett. 48 (1983) 975; {\it
Topologically massive gauge theories}, Ann. Phys. NY 140 (182) 372.



\bibitem{yang} B-H. Lee, K. Lee and H.S. Yang, {\it The CP(n) model on
noncommutative plane} Phys. Lett. B498 (2001) 277, hep-th/0007140.

\bibitem{dunne} G. Dunne, K. Lee and C. Lu, {\it On the finite
temperature Chern-Simons coefficient}, Phys. Rev. Lett. 78 (1997)
3434. 



\bibitem{angular}
R. Jackiw, Phys. Rev. Lett. 41 (1978) 1635.




\bibitem{piljin} K. Lee, Phys. Rev. D49 (1994) 4265; K. Lee and P. Yi,
Phys. Rev. D52 (1995) 2412.

\bibitem{yoonbai} Y. Kim and K. Lee,  Phys. Rev. 49 (1994) 2041.



\bibitem{nair} V.P. Nair and A.P. Polychronakos, {\it On level
quantization for the noncommutative Chern-Simons theory}, hep-th/0102181.





\end{thebibliography}
\end{document}